\documentclass[twocolumn,showpacs,showkeys,nofootinbib,pre]{revtex4}
\usepackage{epsfig}
\usepackage{graphicx}
\usepackage{soul}
\usepackage{color}
\usepackage{amsmath}
\usepackage{stackrel}
\begin{document}
\title{Controlled Generation of Chimera States in SQUID Metasurfaces using DC 
Flux Gradients}
\author{N. Lazarides$^{1,2}$, J. Hizanidis$^{1,2}$, G. P. Tsironis$^{1,2}$}
\affiliation{
$^{1}$Department of Physics, University of Crete, P. O. Box 2208, 
      71003 Heraklion, Greece; \\
$^{2}$National University of Science and Technology ``MISiS'', 
      Leninsky Prospekt 4, Moscow, 119049, Russia
}
\date{\today}
\begin{abstract}
SQUID (Superconducting QUantum Interference Device) metamaterials, subject to a
time-independent (dc) flux gradient and driven by a sinusoidal (ac) flux field, 
support chimera states that can be generated with zero initial conditions. The 
dc flux gradient and the amplitude of the ac flux can control the number of 
desynchronized clusters of such a generated chimera state (i.e., its ``heads'') 
as well as their location and size. The combination of three measures, i.e., the
synchronization parameter averaged over the period of the driving flux, the 
incoherence index, and the chimera index, is used to predict the generation of a 
chimera state and its multiplicity on the parameter plane of the dc flux 
gradient and the ac flux amplitude. Moreover, the full-width half-maximum of the 
distribution of the values of the synchronization parameter averaged over the 
period of the ac driving flux, allows to distinguish chimera states from 
non-chimera, partially synchronized states.
\end{abstract}
\maketitle
\section{Introduction.}
SQUID (Superconducting QUantum Interference Device) metamaterials have been 
investigated intensively both experimentally and theoretically the last decade
\cite{Anlage2011,Jung2014,Lazarides2018b}, revealing a number of extaordinay
properties. From the dynamical systems point of view, SQUID metamaterials are 
extended systems of identically coupled, highly nonlinear oscillators. 
Interestingly, they support counter-intuitive spatiotemporal states such as 
{\em chimera states}, as it was demonstrated numerically both for locally and 
nonlocally coupled SQUIDs in one and two dimensions 
\cite{Lazarides2015b,Hizanidis2016a,Hizanidis2016b,Hizanidis2019}. 
Chimera states have attracted great attention by the scientific community since 
their discovery \cite{Kuramoto2002} from both theoretical and experimental 
viewpoints \cite{Panaggio2015,Scholl2016,Yao2016}. In the present case they are 
characterized by the coexistence of clusters of SQUIDs, in which flux 
oscillations are synchronized and desynchronized.

There have been several attempts to control the emergence and stability of 
chimera states through, e.g., the introduction of excitable units in 
FitzHugh-Nagumo networks \cite{Isele2016}, by feedback in coupled phase 
oscillators \cite{Sieber2014,Semenov2016}, by using spatial pinning in a ring of 
nonlocally coupled oscillators \cite{Gambuzza2016}, or by using gradient 
dynamics in non-locally coupled rings \cite{Bick2015}. Here, the generation and 
control of chimera states in two-dimensional (2D) SQUID metamaterials (SQUID 
metasurfaces) driven by a sinusoidal (ac) flux field and biased by a 
time-independent (dc) flux gradient is demonstrated numerically. Moreover, they 
are generated without the need for a specific choice of initial conditions. Note 
that the application of a dc flux gradient to a SQUID metamaterial is 
experimentally feasible with existing experimental set-ups \cite{Zhang2015}. The 
size, location, and the number of desynchronized clusters (i.e., the 
multiplicity) of the generatd chimera states can be controlled by the dc flux 
gradient and the ac flux amplitude. Such spatially inhomogeneous states can be 
in principle detected and visualized using a recently developed method that 
relies on the cryogenic Laser Scanning Microscope (LSM) 
\cite{Zhuravel2016b,Averkin2016,Karpov2018}.
\begin{figure}[h!]
\includegraphics[angle=0,width=1 \linewidth]{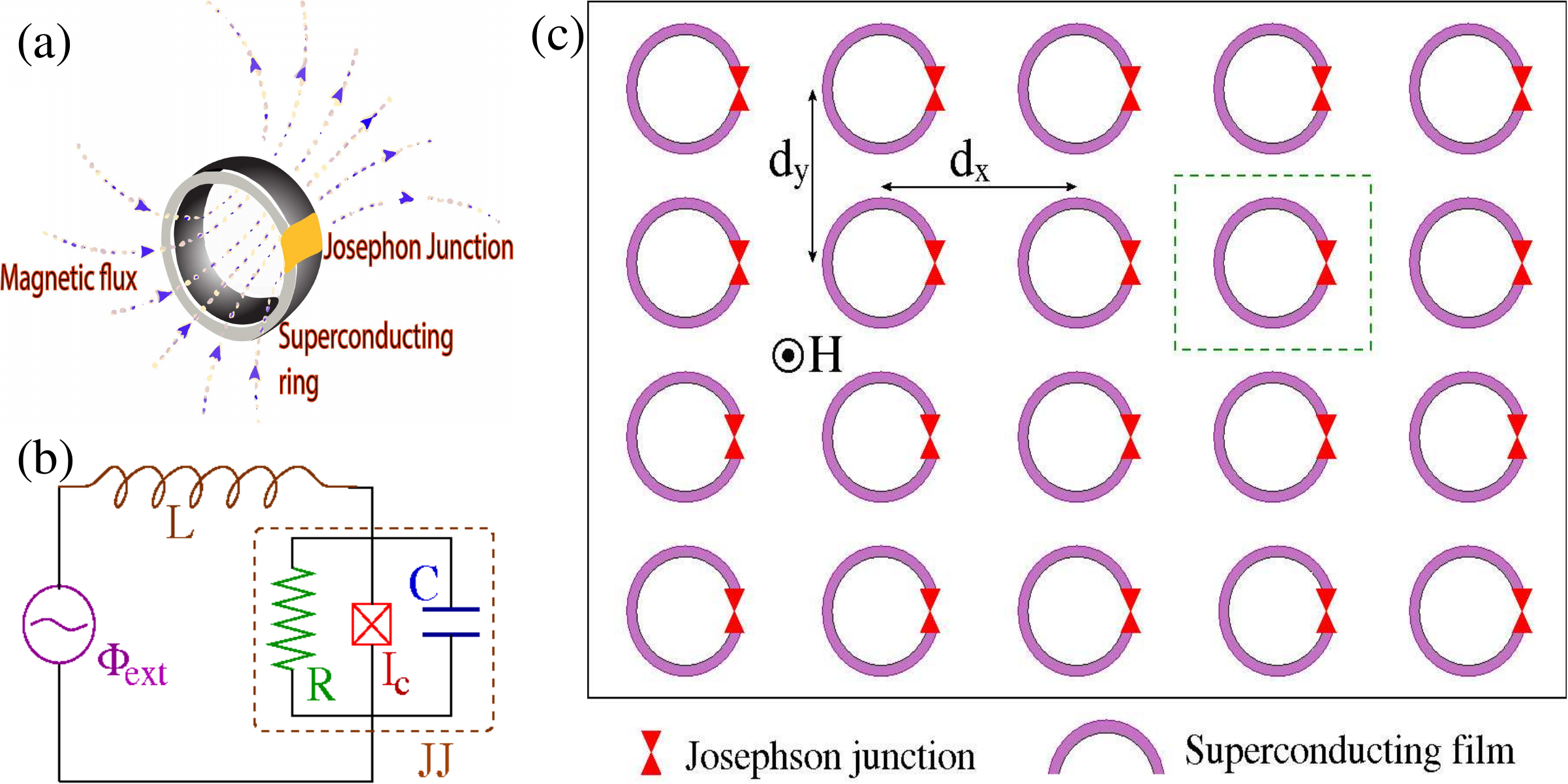}
\caption{(Color online)
(a) Schematic of a superconducting quantum interference device (SQUID) in a 
    magnetic field.   
(b) Equivalent electrical circuit.
(c) Schematic of a two-dimensional SQUID metamaterial (SQUID metasurface).
}
\label{fig01}
\end{figure}

\section{Flux dynamics equations.}
A simple version of a SQUID, consisting of a superconducting ring interrupted 
by a Josephson junction (JJ) \cite{Josephson1962}, is shown schematically in 
Fig. \ref{fig01}(a). Its equivalent electrical circuit model features a 
self-inductance $L$, a capacitance $C$, a resistance $R$, and a critical current 
$I_c$, which characterizes the JJ (Fig. \ref{fig01}(b)). SQUIDs have been 
studied for many years and they have found numerous applications in magnetic 
field sensors, biomagnetism, non-destructive evaluation, and gradiometers, among 
others \cite{Clarke2004a,Clarke2004b}. They exhibit rich dynamic behavior 
including multistability, complex bifurcation structure, and chaos 
\cite{Hizanidis2018}. The multistability property, in particular, favors the 
emergence of chimera states in SQUID metamaterials/metasurfaces 
\cite{Hizanidis2016a,Hizanidis2016b}.

Consider a planar SQUID metamaterial (Fig. \ref{fig01}(c)), in which $N\times N$ 
identical SQUIDs are arranged on a tetragonal lattice ($d_x =d_y$), and they are 
coupled to their nearest-neighbors through magnetic dipole-dipole forces due to 
their mutual inductance $M$. The dynamic equations for the (normalized) flux 
through the ring of the $(n,m)$th SQUID, $\phi_{n,m}$, are given by 
\cite{Lazarides2013b,Tsironis2014b}
\begin{eqnarray}
\label{eq01}
  \ddot{\phi}_{n,m} +\gamma \dot{\phi}_{n,m} +\phi_{n,m}
   +\beta\, \sin( 2 \pi \phi_{n,m} ) =\phi_{n,m}^{eff} (\tau)
\nonumber \\
  +\lambda ( \phi_{n-1,m} +\phi_{n+1,m} +\phi_{n,m-1} +\phi_{n,m+1} ),
\end{eqnarray}	
where $n,m=1,...,N$, and
\begin{eqnarray}
\label{eq02}
\phi_{n,m}^{eff} =\phi_{n,m}^{ext} 
   -\lambda ( \phi_{n-1,m}^{ext} +\phi_{n+1,m}^{ext} 
             +\phi_{n,m-1}^{ext} +\phi_{n,m+1}^{ext} ),
\end{eqnarray}
with
\begin{eqnarray}
\label{eq03}
  \phi_{n,m}^{ext} = \phi_{n,m}^{dc}  +\phi_{ac} \cos(\Omega \tau )
\end{eqnarray}
In Eqs. (\ref{eq01})-(\ref{eq03}), $\lambda =M/L$ is the coupling strengh between 
neighboring SQUIDs, and 
\begin{equation}
\label{eq04}
  \beta =\frac{\beta_L}{2\pi} =\frac{L\, I_c}{\Phi_0}, \qquad
  \gamma= \omega_{LC} \frac{L}{R},
\end{equation}
is the rescaled SQUID parameter and loss coefficient, respectively, with 
$\omega_{LC} = 1/\sqrt{LC}$ being the inductive-capacitive ($L C$) or 
geometrical SQUID frequency ($\Phi_0$ is the flux quantum). The overdots on 
$\phi_{n,m}$ denote differentiation with respect to the temporal variable $\tau$ 
(normalized to $\omega_{LC}^{-1}$). For applying a dc flux gradient along the 
direction of increasing $n$, a dc flux function of the form 
\begin{equation}
\label{eq06}
  \phi_{n,m}^{dc} =\frac{n-1}{N-1} \phi_{dc}^{max},
\end{equation}
where $n,m=1, ... ,N$ is assumed. Thus, the dc flux increases linearly from zero 
(for the SQUIDs at $(n,m)=(1,m)$, $m=1,...,N$) to $\phi_{dc}^{max}$ (for the 
SQUIDs at $(n,m)=(N,m)$, $m=1,...,N$). For the numerical simulations, the 
experimentally relevant model parameters $\beta_L =0.86$, $\gamma =0.01$, and 
$\lambda =-0.02$, have been adopted \cite{Trepanier2013}. The driving frequency 
has been chosen to be close to the geometrical resonance, i.e., $\Omega =1.01$,
where individual SQUIDs exhibit extreme multistability \cite{Hizanidis2018}.

\section{Generation and Control of Chimera States.}
\begin{figure*}[t!]
\includegraphics[angle=0,width=0.9 \linewidth]{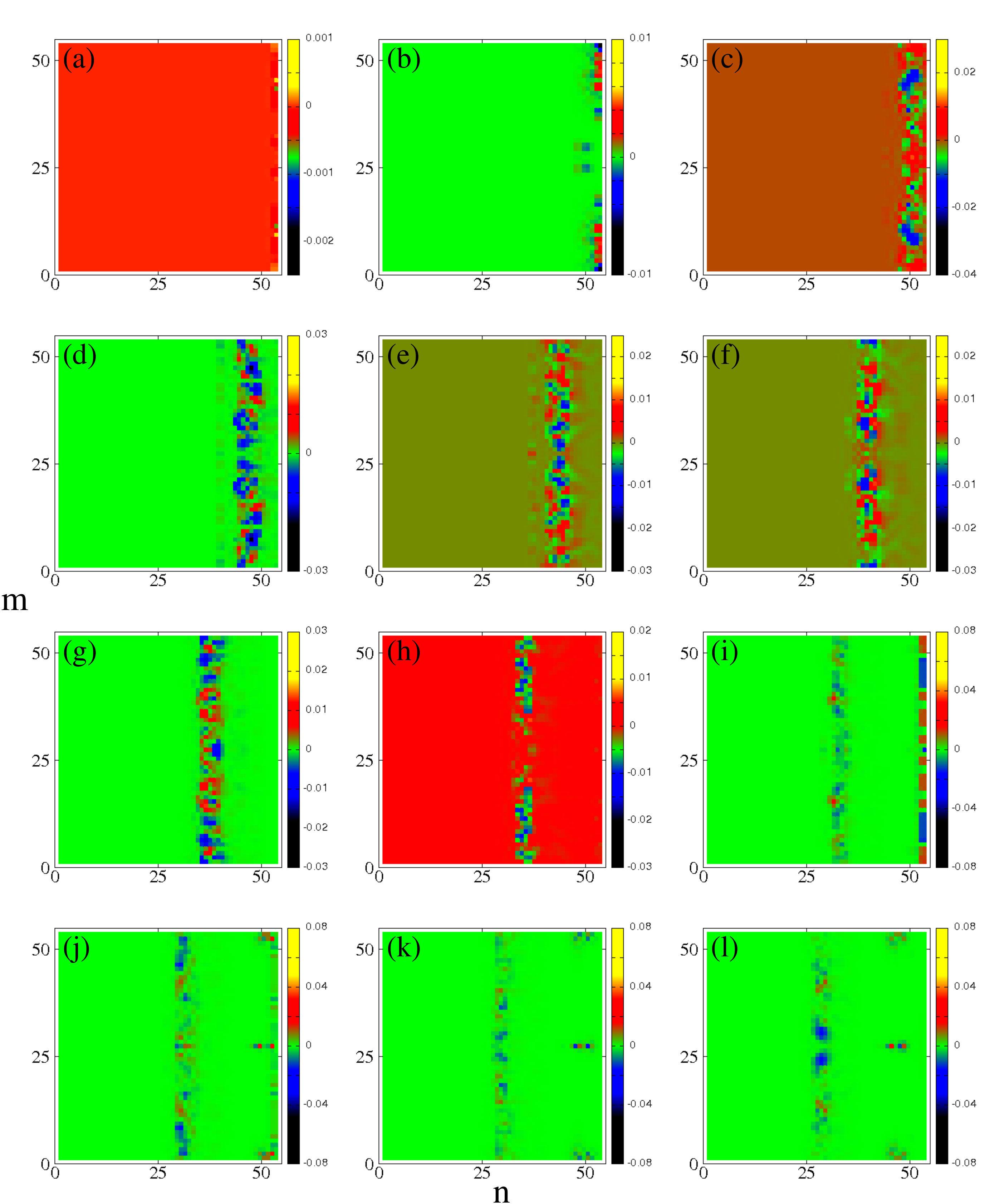}
\caption{(Color online)
 Maps of $\left< \dot{\phi}_{n,m} \right>_T$ on the $n-m$ plane for $N=54$, 
 $\beta_L =0.86$, $\gamma=0.01$, $\lambda=-0.02$, $\Omega=1.01$, 
 $\phi_{ac} =0.02$, and $\phi_{dc}^{max}$ from (a) to (l) increases from $0.27$ 
 to $0.60$ in steps of $0.03$. 
}
\label{fig02}
\end{figure*}

\begin{figure*}[t!]
\includegraphics[angle=0,width=0.9 \linewidth]{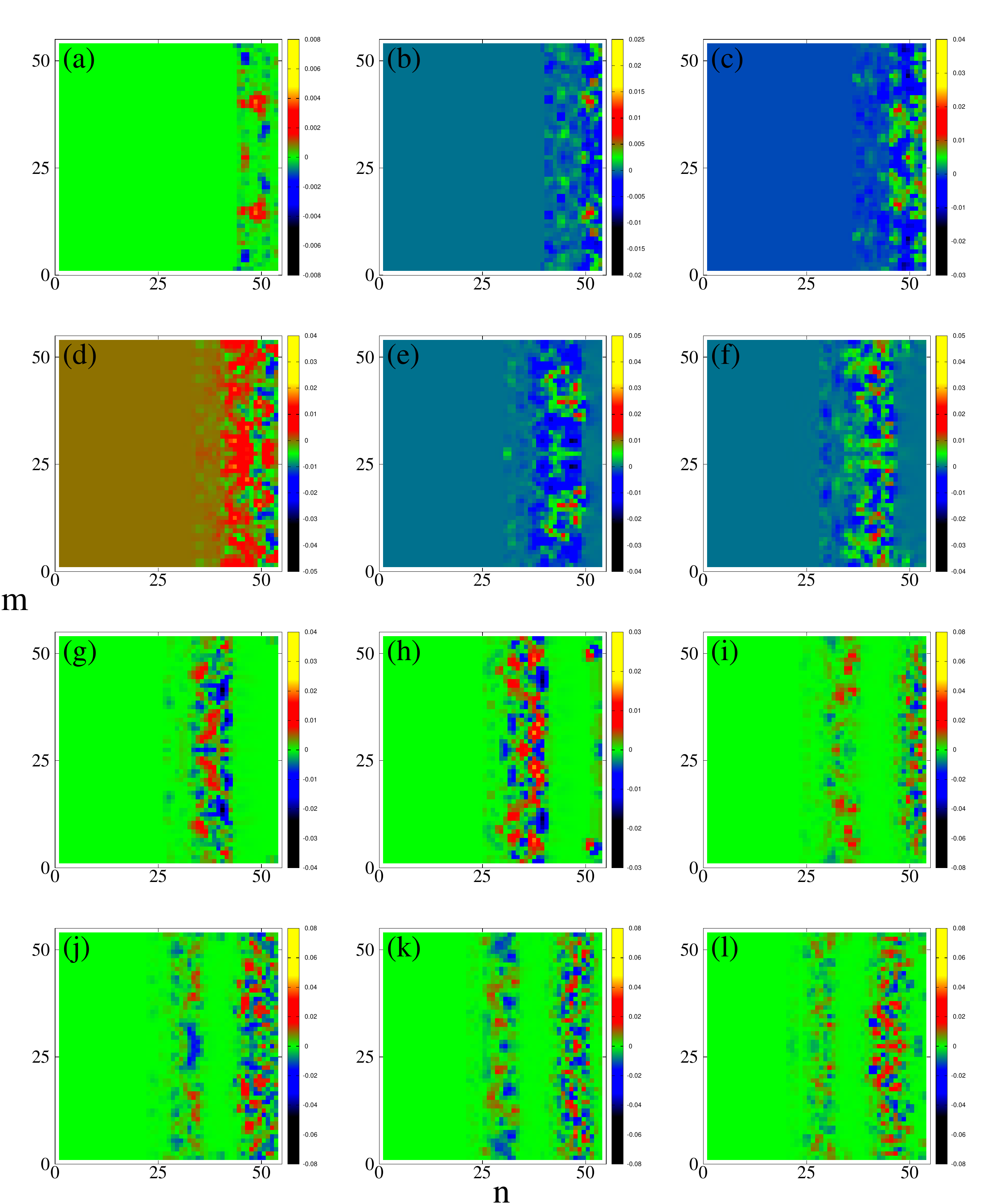}
\caption{(Color online)
 Maps of $\left< \dot{\phi}_{n,m} \right>_T$ on the $n-m$ plane for 
 $\phi_{ac} =0.04$ and the other parameters as in Fig. \ref{fig02}.
}
\label{fig03}
\end{figure*}

Eqs. (\ref{eq01}) are integrated numerically in time using a standard 
fourth-order Runge-Kutta algorithm with time-step $h=0.02$ and free-end boundary 
conditions. For any parameter set, Eqs. (\ref{eq01}) are initialized with zeros, 
i.e., with 
\begin{equation}
\label{eq07.2}
   \phi_{n,m} (\tau=0) =0,~~ \dot{\phi}_{n,m} (\tau=0) =0, 
   ~~{\rm for~ any}~ n, m. 
\end{equation}
Time-integration for $10^4 ~T$ time-units, where $T=2\pi/\Omega$ is the driving 
period, is allowed for transients to die-out and a ``steady-state'' to be 
reached. In Figs. \ref{fig02} and \ref{fig03}, the averages of 
$\dot{\phi}_{n,m} (\tau)$ over a driving period $T$, i.e.,  
\begin{equation}
\label{eq08}
  \left<\dot{\phi}_{n,m} \right>_T =
   \frac{1}{T} \int_0^T \dot{\phi}_{n,m} \, d\tau,
\end{equation}
are mapped on the $n-m$ plane for $\phi_{ac} =0.02$ and $\phi_{ac} =0.04$,
respectively, and several values of $\phi_{dc}^{max}$ (which determines the 
value of the gradient of the dc flux). In these maps, areas of uniform (resp. 
nonuniform) colorization indicate that the SQUID oscillators there are 
synchronized (resp. desynchronized).

For low $\phi_{dc}^{max}$, the SQUID metasurface is in an almost spatially 
homogeneous state. With increasing $\phi_{dc}^{max}$ (or equivalently by 
increasing flux gradient), a threshold is reached at which some of the SQUIDs 
close to the $n=N$ end become desynchronized with respect to the others, and the 
spatially homogeneous state breaks down. In Fig. \ref{fig02}, the breaking of 
the spatially homogeneous state of the SQUID metasurface occurs at 
$\phi_{dc}^{max} =0.27$ (Fig. \ref{fig02}(a)). With further increase of 
$\phi_{dc}^{max}$, more and more SQUIDs close to $n=N$ become desynchronized 
(Fig. \ref{fig02}(b)) until they form a well-defined desynchronized (incoherent) 
cluster at $\phi_{dc}^{max} =0.33$ (Fig. \ref{fig02}(c)). That incoherent 
cluster begins to shift towards the $n=1$ end with $\phi_{dc}^{max}$ increasing 
yet further (Fig. \ref{fig02}(d) - \ref{fig02}(g)), i.e., from 
$\phi_{dc}^{max} =0.33$ to $\phi_{dc}^{max} =0.45$. For values of 
$\phi_{dc}^{max}$ in this interval, the width of the incoherent cluster remains 
approximately the same. In Figs. \ref{fig02}(h) - \ref{fig02}(l), i.e., from 
$\phi_{dc}^{max} =0.48$ to $\phi_{dc}^{max} =0.60$, the incoherent cluster 
continues to shift towards the $n=1$ end but its width has become narrower. In 
Fig. \ref{fig02}(i) ($\phi_{dc}^{max} =0.51$), a second incoherent cluster 
emerges at $n \sim N$ which also changes with further increasing 
$\phi_{dc}^{max}$. This effect becomes more clear in Fig. \ref{fig03}, whose 
parameters differ from those of Fig. \ref{fig02} only in the value of 
$\phi_{ac}$. 
In Fig. \ref{fig03}, the threshold value of $\phi_{dc}^{max}$ for 
desynchronization to appear at the $n=N$ end is lower. Indeed, as can be 
observed clearly in Fig. \ref{fig03}(a), for $\phi_{dc}^{max} =0.27$, the 
desynchronized cluster is already formed close to the $n=N$ end (compare with 
the corresponding Fig. \ref{fig02}(a)). Similarly to what is observed in Fig. 
\ref{fig02}, the desynchronized cluster shifts with increasing $\phi_{dc}^{max}$ 
towards the $n=1$ end. Its width remains approximately the same from 
$\phi_{dc}^{max} =0.30$ to $\phi_{dc}^{max} =0.48$ 
(Fig. \ref{fig03}(b) - \ref{fig03}(h)). From $\phi_{dc}^{max} =0.51$ to 
$\phi_{dc}^{max} =0.60$ (Fig. \ref{fig03}(i) - \ref{fig03}(l)) the 
desynchronized cluster continues its shifting towards the $n=1$ end, while its 
width becomes somewhat narrower. Here, the formation and shifting of a second 
desynchronized cluster can be clearly observed (due to higher ac flux amplitude
$\phi_{ac} =0.04$). The second desynchronized cluster appears at the $n=N$ 
end in Fig. \ref{fig03}(h) ($\phi_{dc}^{max} =0.60$), then grows in width and 
follows the shifting pattern of the first desynchronized cluster. Its width 
remains approximately the same for $\phi_{dc}^{max} =0.54 - 0.60$ 
(Figs. \ref{fig03}(j) - \ref{fig03}(l)). The emergence of more desynchronized 
clusters can be observed in larger SQUID metasurfaces, i.e., with larger $N$, 
for sufficiently high values of $\phi_{dc}^{max}$ (not shown).

\section{Characterization of Spatially Inhomogeneous States.}
Eqs. (\ref{eq01}) are integrated in time as above for many values of $\phi_{ac}$ 
and $\phi_{dc}^{max}$ on the $\phi_{ac} - \phi_{dc}^{max}$ plane. Here, after 
the transients have died out, Eqs. (\ref{eq01}) are integrated for 
$\tau_{sst} =10^4$ more time-units, while the SQUID metamaterial is in a steady 
state. The kind of state that has emerged for each pair of $\phi_{ac}$ and 
$\phi_{dc}^{max}$ values can be inferred by calculating several measures which 
are mapped on the $\phi_{ac} - \phi_{dc}^{max}$ plane. A global measure of 
synchronization of the SQUID metasurface is given by 
\begin{equation}
\label{eq09}
   \left< r \right>_{sst} 
   =\frac{1}{\tau_{sst}} \int_0^{\tau_{sst}} r (\tau)\, d\tau, 
\end{equation} 
i.e., by the average over the steady-state integration time $\tau_{sst}$ of the 
magnitude of the Kuramoto synchronization parameter 
$r (\tau) =\left| \Psi (\tau)  \right|$, where
\begin{equation}
\label{eq10}
  \Psi (\tau)=\frac{1}{N^2} \sum_{n,m} e^{2\pi i \phi_{n,m} (\tau)}. 
\end{equation} 
Also, a measure of incoherence and a chimera index for the resulting state can 
be calculated as follows \cite{Gopal2014,Gopal2018}. First, define 
\begin{equation}
\label{eq10.2}
   u_n (\tau) 
   =\left< \frac{1}{N} \sum_{m=1}^{N} \dot{\phi}_{n,m} \right>_T (\tau)
\end{equation} 
where the angular brackets indicate averaging over the driving period $T$, and 
its local spatial average in a region of length $n_0+1$ around the site $n$ at 
time $\tau$, 
\begin{equation}
\label{eq10.3}
   \bar{u}_n (\tau) =\frac{1}{n_0 +1} \sum_{n=-n_0/2}^{+n_0/2} u_n (\tau), 
\end{equation} 
where $n_0 <N$ is an integer. Then, the local standard deviation of $u_n (\tau)$ 
is defined as 
\begin{equation}
\label{eq11}
   \sigma_n (\tau) \equiv \left< \sqrt{ \frac{1}{n_0 +1}
   \sum_{n=-n_0/2}^{+n_0/2} \left( u_n -\bar{u}_n \right)^2 } \right>_{sst},
\end{equation} 
where the large angular brackets denote averaging over $\tau_{sst}$. The index 
of incoherence and the chimera index is then defined as
\begin{equation}
\label{eq12}
   S=1 -\frac{1}{N} \sum_{n=1}^N s_n, \qquad 
        \eta =\frac{1}{2}\sum_{n=1}^N |s_n -s_{n+1}|, 
\end{equation} 
respectively, where $s_n=\Theta(\delta -\sigma_n)$ with $\Theta$ being the Theta 
function. The index $S$ takes its values in $[0,1]$, with $0$ and $1$ 
corresponding to synchronized and desynchronized states, respectively, while all 
other values between them indicate a chimera or multi-chimera state. The index 
$\eta$ equals to unity (an integer greater than unity) for a chimera 
(a multi-chimera) state. It roughly provides the number of desynchronized 
clusters or ``heads'' of the (multi-)chimera state.
\begin{figure}[h!]
\includegraphics[angle=0,width=1 \linewidth]{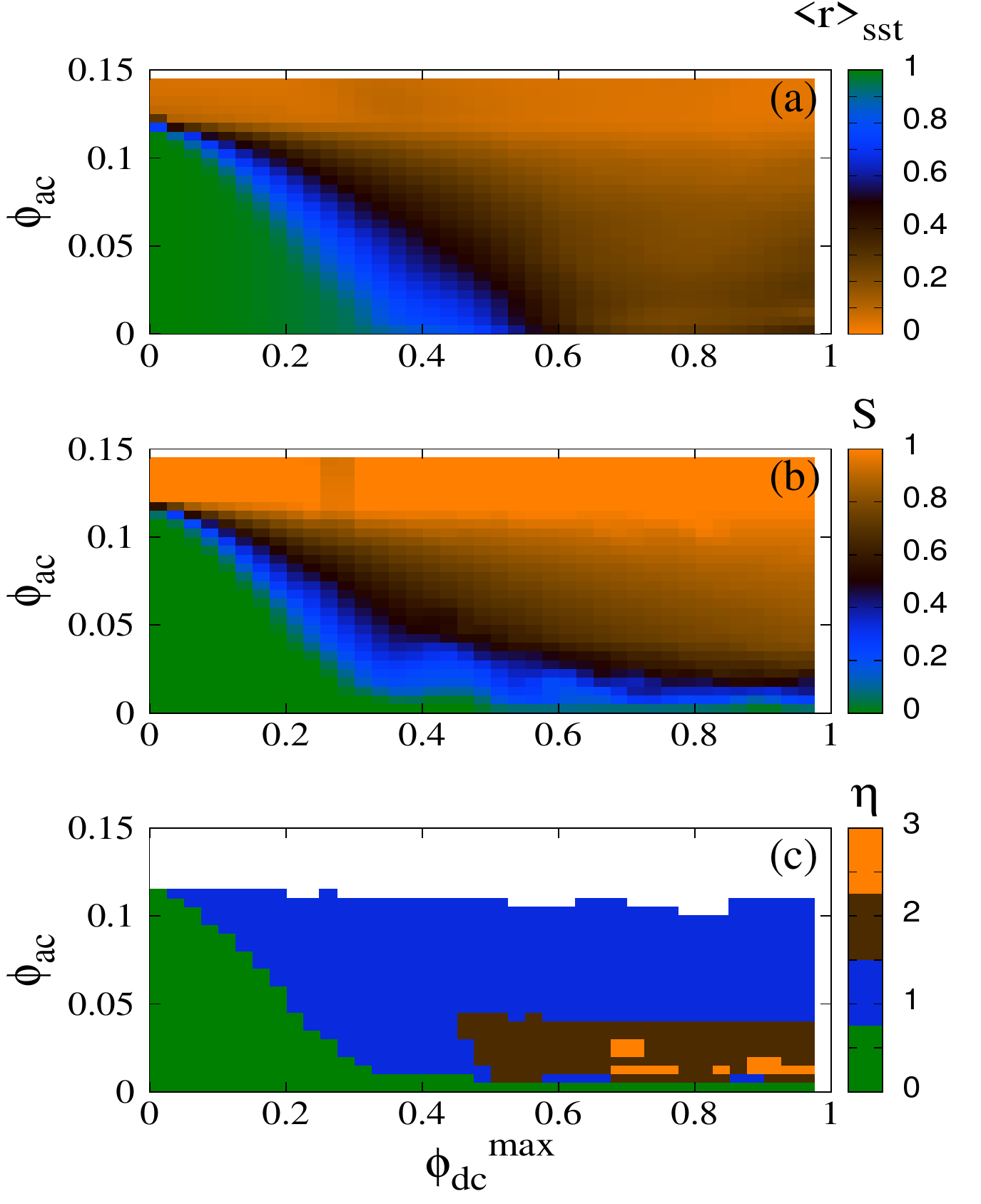}
\caption{(Color online)
 Maps of 
 (a) the magnitude of the synchronization parameter averaged over the 
     steady-state integration time $\tau_{sst}$, $\left< r \right>_{sst}$, 
 (b) the incoherence index $S$, and 
 (c) the chimera index $\eta$, on the $\phi_{dc}^{max} - \phi_{ac}$ plane for 
     $N=54$, $\beta_L =0.86$, $\gamma=0.01$, $\lambda =-0.02$, and 
     $\Omega \simeq 1.01$.
}
\label{fig04}
\end{figure}

In Fig. \ref{fig04}, the measures $\left< r \right>_{sst}$, $S$, and $\eta$,
calculated from Eqs. (\ref{eq09}) and (\ref{eq12}), are mapped on the 
$\phi_{dc}^{max} - \phi_{ac}$ plane. In Fig. \ref{fig04}(a), for values of 
$\phi_{dc}^{max}$ and $\phi_{ac}$ within the green area, in which 
$\left< r \right>_{sst} \simeq 1$, the SQUID metamaterial is in a state with a 
high degree of synchronization (almost synchronized state). In the rest of the 
plane, $\left< r \right>_{sst}$ takes values significantly less than unity, 
indicating that the SQUID metamaterial is in a partially synchronized or in a 
completely desynchronized state. However, it cannot be concluded from 
$\left< r \right>_{sst}$ alone whether a particular state is partially or 
completely desynchronized, and furthermore it cannot be concluded whether a 
partially synchronized state is a chimera or multi-chimera state or another type 
of spatially inhomogeneous state. For that purpose, the information in Figs. 
\ref{fig04}(b) and (c) for the the incoherence index $S$ and the chimera index 
$\eta$ has to be used. In Fig. \ref{fig04}(b), the incoherence index $S$ is 
exactly zero in the green area ($S=0$), indicating synchronization. It should be 
noted however that the green areas in Figs. \ref{fig04}(a) and (b) do not 
completely coincide. Indeed, in Fig. \ref{fig04}(b), the green area 
extends to $\phi_{dc}^{max}$ values up to $0.5$ (for very low $\phi_{ac}$),
for which Fig. \ref{fig04}(a) gives $\left< r \right>_{sst}$ significantly less
than unity. This apparent contradiction is discussed in the next paragraph. The 
chimera index $\eta$, mapped in Fig. \ref{fig04}(c), provides information for 
the number of ``heads'' of a chimera/multi-chimera state, i.e., the number of 
desynchronized clusters. In the green area, that number is zero as it should be, 
while it is one, two, and three in the blue, brown, and orange areas, indicating 
the generation of single-headed, two-headed, and three-headed chimera states,
respectively. Note also the white area in Fig. \ref{fig04}(c), for 
$\phi_{ac} \gtrsim 0.11$, for which the incoherence index $S$ in Fig. 
\ref{fig04}(b) is exactly unity. In this area, the resulting state of the SQUID 
metamaterial is completely desynchronized.
\begin{figure}[h!]
\includegraphics[angle=0,width=1 \linewidth]{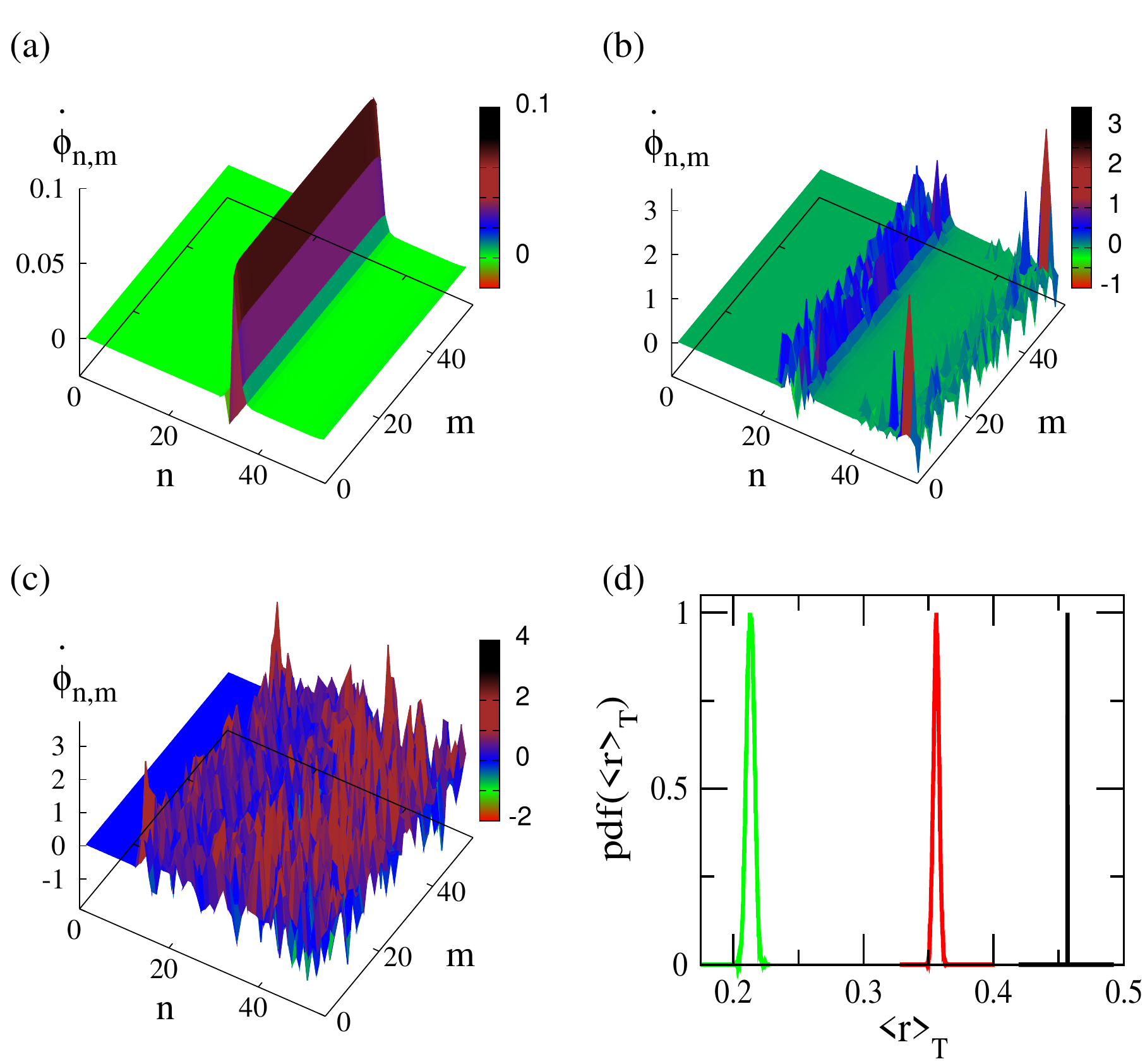}
\caption{(Color online)
 Two-dimensional profiles of $\dot{\phi}_{n,m}$ for $\phi_{dc}^{max} =0.6$, 
 $\beta_L =0.86$, $\gamma=0.01$, $\lambda =-0.02$, $\Omega \simeq 1.010$$N=54$,
 and
 (a) $\phi_{ac}=0.001$,
 (b) $\phi_{ac}=0.03$, 
 (c) $\phi_{ac}=0.08$.
 (d) Probability density functions $pdf(\left< r \right>_{T})$ for the states
 in (a), (b), and (c), shown as black, red, and green curves, respectively. 
}
\label{fig05}
\end{figure}

As mentioned in the previous paragraph, there is a green narrow strip in Figs. 
\ref{fig04}(b) and (c) from $\phi_{dc}^{max} \simeq 0.32$ to $0.98$ and very low
values of $\phi_{ac}$, in which $S=0$ and $\eta=0$ that indicate a non-chimeric 
state. At the same time, the measure $\left< r \right>_{sst}$ in the same area
assumes values significantly lower than unity 
($\left< r \right>_{sst} \simeq 0.46$), indicating at least partial 
desynchronization. 
For clarifying this apparent descrepancy, three spatial profiles are presented 
in Fig. \ref{fig05}(a), (b), and (c), which are obtained for 
$\phi_{dc}^{max} =0.6$ and $\phi_{ac}=0.001$, $0.03$, and $0.08$, respectively. 
The corresponding pairs of values are located in the green strip, the blue area
and the brown area, respectively, of Fig. \ref{fig04}(b). The profile in Fig. 
\ref{fig05}(a), for $\phi_{ac}=0.001$, is clearly not a chimera state; however, 
it is not a synchronized state either, since the fluxes in at least two or three 
SQUIDs oscillate with a high amplitude compared to the fluxes in the rest of 
them (and also their phases with respect to the driver are generally different). 
This justifies the low value of $\left< r \right>_{sst}$ in Fig. \ref{fig04}(a)
(at least partial desynchronization), as well as the value $S=0$ (no chimera 
state) and $\eta=0$ (no desynchronized cluster). In Fig. \ref{fig05}(b), 
for $\phi_{ac}=0.03$, a two-headed chimera state is observed, for which 
$\left< r \right>_{sst} \simeq 0.36$, $S=0.48$, and $\eta =2$. In Fig. 
\ref{fig05}(c), a single-headed chimera state with 
$\left< r \right>_{sst} \simeq 0.21$, $S=0.81$, and $\eta =1$ is obtained. Thus, 
although the measures $S$ and $\eta$ correctly predict the existence of a 
chimera state and its ``heads'', they cannot discriminate between a synchronized
state and a clustered state in which the oscillators in each cluster are 
synchronized, but the clusters are not synchronized to each other. On the other 
hand, in such a clustered state, $\left< r \right>_{sst} < 1$ indicating at 
least partial desynchronization. In order to distinguish between such clustered 
states and synchronized states without inspecting the profiles, additional 
information is necessary, which can be provided, e.g., by the probability 
distribution function of the $\left< r \right>_{T}$ values, 
$pdf(\left< r \right>_{T})$, where
\begin{equation}
\label{eq13}
   \left< r \right>_T =\frac{1}{T} \int_0^{T} r (\tau)\, d\tau. 
\end{equation} 
The distributions for the three profiles in Figs. \ref{fig05}(a)-(c) are shown 
in Fig. \ref{fig05}(d). The full-width half-maximum (FWHM) of the distributions 
is a measure of metastability for chimera states \cite{Shanahan2010,Wildie2012}. 
However, for states such as that shown in Fig. \ref{fig05}(a), for which 
$\left< r \right>_{T}$ does not fluctuate in time, the FWHM of the 
$pdf(\left< r \right>_{T})$ is practically zero (black curve). For the two 
chimera states in Figs. \ref{fig05}(b) and (c), the corresponding FWHM is 
$\sim 0.005$ and $\sim 0.007$, respectively. It should be noted that there are 
also other measures that could be employed such as, e.g., a measure based on the 
local curvature of a given state \cite{Kemeth2016}. This measure has been proved 
particularly useful whenever turbulent chimeras appear, e.g., in semiconductor 
laser arrays \cite{Shena2017a,Shena2017b}.

\section{Concluding.}
In the present work it is demonstrated that chimera states can be generated and
controlled by a dc flux gradient. Note that the application of a dc flux gradient 
on a SQUID metamaterial is experimentally feasible with existing experimental 
set-ups \cite{Zhang2015}. It is importnt that by using a dc flux gradient, there 
is no need for any specific initialization of the SQUID metamaterial to obtain 
chimera states. The appearance (i.e., the location and size of the 
desynchronized cluster(s), the number of ``heads'' of the chimera state) and the 
degree of synchronization of the chimera states can be controlled to some extent 
by varying the externally controlled parameters such as the dc flux gradient 
(determined by $\phi_{dc}^{max}$), and the ac flux amplitude $\phi_{ac}$.

Three measures, i.e., $\left< r \right>_{sst}$, $S$, and $\eta$, are calculated 
and mapped on the $\phi_{dc}^{max} - \phi_{ac}$ plane. By combining information 
from all these three measures, the generation or not of a chimera state, and the 
number of its heads can be anticipated. However, neither of these measures can 
distinguish between a chimera state and a clustered state in which the SQUIDs
within each clusters are syncronized but the clusters are not synchronized to 
each other. This can be however achieved by calculating an additional quantity,
the FWHM of the probability density $pdf(\left< r \right>_{T})$; chimera states 
are then indicated by nonzero FWHM. 

\section*{ACKNOWLEDGMENT}
The authors gratefully acknowledge the financial support of the Ministry of 
Science and Higher Education of the Russian Federation in the framework of 
Increase Competitiveness Program of NUST ``MISiS'' (No. K2-2017-006), 
implemented by a governmental decree dated 16th of March 2013, N 211. 
This research has been financially supported by 
General Secretariat for Research and Technology (GSRT) and the Hellenic 
Foundation for Research and Innovation (HFRI) (Grant No. 203).



\end{document}